Resilience of Urban Transport Network-of-Networks under Intense Flood Hazards Exacerbated by Targeted Attacks

**Nishant Yadav**[1], **Samrat Chatterjee**[2,†], **Auroop R. Ganguly**[1,†]

[1]Sustainability and Data Sciences Laboratory, Department of Civil and Environmental Engineering, Northeastern University, Boston, MA, USA
[2]Computing and Analytics Division, National Security Directorate, Pacific Northwest National Laboratory, Richland, WA, USA
[†]Corresponding Authors: a.ganguly@northeastern.edu; samrat.chatterjee@pnnl.gov

**Abstract**
Natural hazards including floods can trigger catastrophic failures in interdependent urban transport network-of-networks (NoNs). Population growth has enhanced transportation demand while urbanization and climate change have intensified urban floods. However, despite the clear need to develop actionable insights for improving the resilience of critical urban lifelines, the theory and methods remain underdeveloped. Furthermore, as infrastructure systems become more intelligent, security experts point to the growing threat of targeted cyber-physical attacks during natural hazards. Here we develop a hypothesis-driven resilience framework for urban transport NoNs, which we demonstrate on the London Rail Network (LRN). We find that topological attributes designed for maximizing efficiency rather than robustness render the network more vulnerable to compound natural-targeted threats including cascading failures. Our results suggest that an organizing principle for post-disruption recovery may be developed with network science principles. Our findings and frameworks can generalize to urban lifelines and more generally to real-world spatial networks.

1. Introduction

According to the World Economic Forum's Global Risks Report 2019 [1], extreme weather events are the global risks of highest concern. Heavy precipitation, along with associated flooding in urban megaregions, has been on the rise both in intensity and frequency under the dual forcings of climate change and rapid urbanization. Urban centers are the economic nerve centers of modern economies, with many densely populated megaregions experiencing further population growth. Thus, critical urban lifeline infrastructure systems (CULIS) across the globe are under stress, with multimodal urban transport systems (MUTS) among the worst affected by urban flooding. Moreover, transportation networks are functionally interdependent with each other and on other infrastructure systems such as the power grid and communication networks. Thus, even a limited disruption in one system can spiral out of control leading to severe loss of lifeline functions. Further, as MUTS are becoming increasingly connected and autonomous, security experts have pointed to the growing threat of opportunistically targeted cyber-attacks designed to take advantage of natural hazard events [2].

Numerous definitions of resilience have been proposed in the literature [3], although here we adopt the most widely cited provided by the US National Academy of Sciences: *"the ability to prepare and plan for, absorb, recovery from and more successfully adapt to adverse events"*. The growing threat of natural, targeted and compound threats on MUTS calls for an urgent need to analyze and build resilience at a system level. Furthermore, the multiscale and interconnected nature of MUTS, combined with the inherent unpredictability of extreme weather events and compound extremes, make the resilience task even more challenging [4]. In addition, resilience of networked systems is often a function of the topology and dynamics, but quantification and modeling the latter remain a challenge for complex systems.

Conceptual frameworks for resilience are available in the extant literature [4, 5], but limited work has been done on modeling and quantifying MUTS resilience with the aim of generating actionable insights for stakeholders. The recently demonstrated "universality" of network science-based approaches [6-10] provide a natural method of choice for quantifying resilience of networked systems such as the MUTS. One of the most widely studied property in network science is the robustness of a network given the failure of a subset of its nodes. Inspired by percolation theory, the giant (largest) connected component of the network is typically treated as a proxy for the state of functionality in the network [11]. This approach has helped in understanding the robustness properties of different network topologies and the corresponding systems they represent. Recent studies in this area range from robustness analysis of specific infrastructure systems such as the power grid [12] and MUTS [9], to "universal" theories of resilience [13, 14] which consider network dynamics and topology. Besides robustness, researchers have applied network science and engineering to find optimal attacker strategies [15] as well as most effective post-failure recovery sequences [16] in infrastructure networks.

While prior studies in transportation networks have primarily focused on single networks (and single disruptions) [9, 33], the reality is that real-world infrastructure networks rarely appear in isolation. The interconnected and interdependent network-of-networks (NoNs) give rise to a rich topology, which in turn exhibits behavior that may be different from single-layer networks. A recent study [17] presented an analytical framework to study the robustness in a system of two interdependent networks and found that such interdependence makes them vulnerable (or, less robust) compared to single networks. The subsequent literature developed generalized results for 'n' interdependent networks [18], diverse failure schemes [19], as well as what has been referred to as "universal theories" for cascading failures in single [13] and interdependent networks-of-networks [14].

The existing literature cited above has largely looked at idealized NoNs from a theoretical standpoint to characterize their physical properties such as percolation threshold and phase transitions. However, theoretical frameworks may not directly apply to real-world networks having topologies which are, often markedly, different. Thus, recent research in real spatial networks (e.g. MUTS) [20], which have not received much attention in the network science literature, reveal that spatial constraints render them significantly more vulnerable compared to their non-embedded counterparts. In addition, prior research has predominantly considered random, natural, and targeted failures individually, however as mentioned earlier, there is a growing need to study compound failure scenarios arising from natural hazards and cyber-physical attacks. Compound failures in this context refer to both simultaneous hazards and attacks, as well as when they occur in sequence, but when the successive events occur before the system has had a chance to recover from the antecedent events.

From the network-science perspective, a long-term goal may be to arrive at universal theories for resilience in spatial network-of-networks. However, given the complexity and diversity among network topologies and failure scenarios, as well as the size of the networks, the possibility of arriving at such universal theories may need to be examined through hypothesis-driven studies. Meanwhile, urgent solutions are needed for such networks; thus, in our opinion a first step would be hypothesis-driven research focusing on specific aspects of the overall problem. Here we address two hypotheses, first that *MUTS NoN topology makes them particularly vulnerable* and second, *even a relatively focused targeted attack, when carried out in conjunction with an intense natural hazard, may cause disproportionate network failure compared to single hazards*.

The primary network-science contribution of this work is to present a generalized framework for a quantitative understanding of resilience, including robustness and recovery, of real-world and spatially constrained urban transportation NoNs. Second, we demonstrate the framework on the London Rail Network and obtain insights that may generalize to other CULIS systems globally. Our analysis has been performed in two parts. First, we focus on understanding the inherent vulnerabilities of the network due to spatial constraints and network sparsity based on current design emphasis on maximizing efficiency. Second, the NoN is tested against a suite of failure scenarios – random, targeted attacks, and natural hazards – as well as the failures owing to compound threats. The insights derived are expected to generalize to other MUTS datasets while the caveats and open challenges may lead to new hypotheses which can be further tested on MUTS datasets globally.

## 2. Results

In the case of 2 layer NoN topology, an analytical solution for robustness of a network (with N nodes) can be codified by the following equation:

$$F(n) = 1 - \lambda \sum_{n=1}^{N} A_{ij} * W(n) * C_{1,2} \qquad [1]$$

where, $F(n)$ is the dynamic network functionality as nodes are removed from the network. $A_{ij}$ is the adjacency matrix which encodes all the information about the network. $W(n)$ represents failure (e.g. random, floods etc.) at each discrete step when node $n$ is removed. $C_{1,2}$ accounts for dependency between the two networks leading to cascading failure. It is to be noted that $A_{ij}$ and $C_{1,2}$ are updated at each failure step. $\lambda$ is the normalizing constant.

In case of compound threats, when a second disruption $T(n)$ occurs after a fraction of nodes $N - N'$ are removed due to the initial failure $W(n)$. Starting from $N'$ the third term in equation below caters to the second failure $T(n)$ (a targeted attack in this case). $A'_{ij}$ and $C'_{1,2}$ is the updated adjacency matrix and dependency at the last instance of initial failure. $\beta$ is the normalizing constant.

$$F(n) = 1 - \lambda \sum_{n=1}^{N-N'} A_{ij} * W(n) * C_{1,2} - \beta \sum_{n=N-N'}^{N} A'_{ij} * T(n) * C'_{1,2} \qquad [2]$$

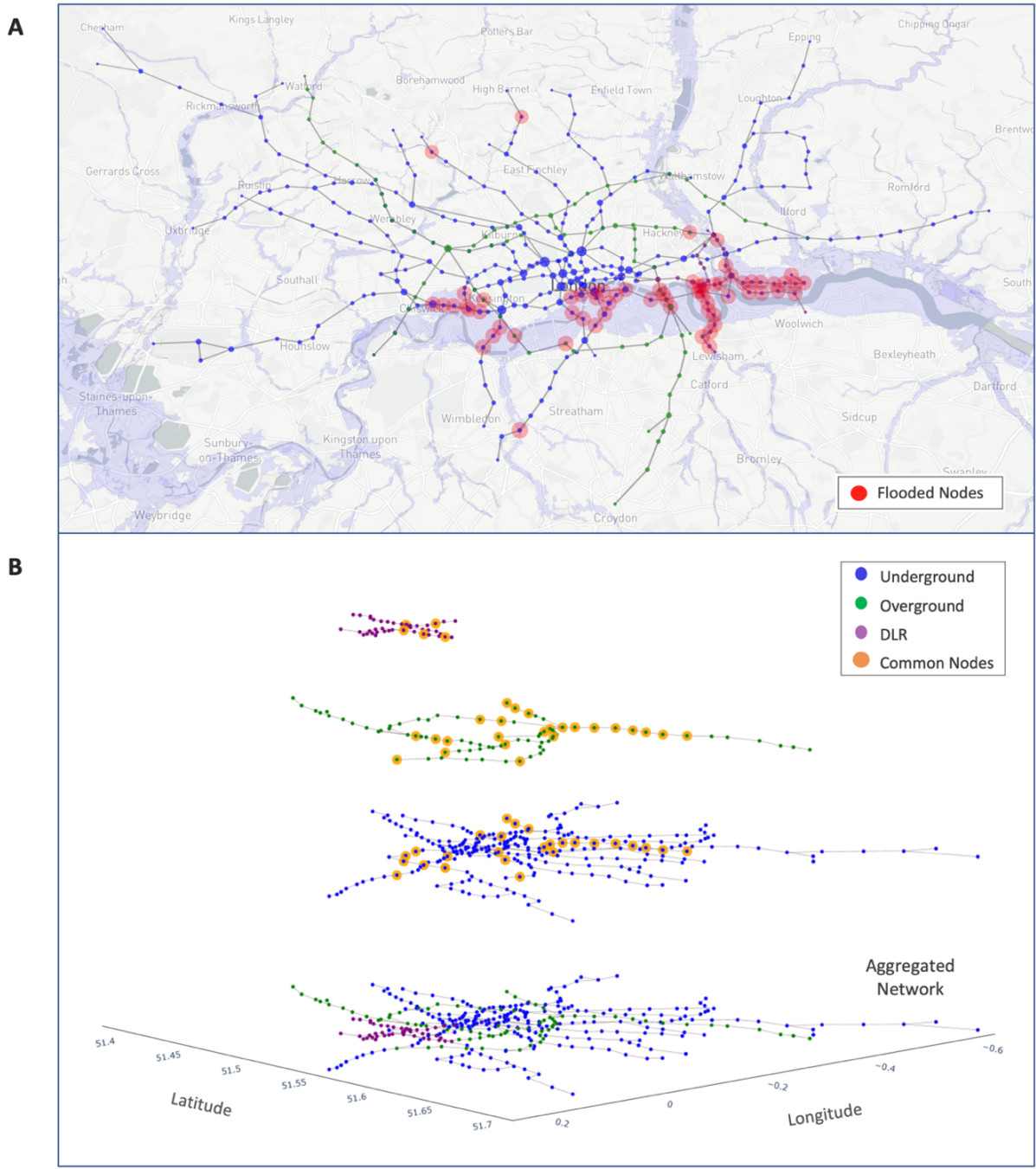

**Figure 1 | Flood on London Rail Network-of-Networks.**
Geocoded network over the map of London along with the 100-1000-year flood risk map (in purple) of river Thames and its tributaries. Nodes in red are the impacted nodes. **(A)** Top view - impacted nodes across all networks (total = 399, impacted = 65) **(B)** Multilayer view - impacted nodes in each network – 31 (Underground), 7 (Overground) and 27 (DLR). Nodes in orange are the shared (common) nodes across the three networks.

## 2.1 Network Vulnerability

The London Rail 'Network-of-Networks' (LRN) comprises of three urban rail networks – Underground, Overground, and the Dockland Light Rail (DLR) – interconnected via shared nodes (see Methods: Common Stations). Although the Underground sub-network itself comprises of 11 different lines, in this work all the Underground lines are considered as a single network. For details on nodes and links in each network, see Methods: Network Structure. Figure 1 (A) shows the geocoded LRN over the map of London. Nodes in red indicate nodes that would be flooded by 100-1000-year floods on the river Thames (relevant details are discussed later).

Robustness of networks, including transportation networks, have been examined in terms of the relative size of the giant (largest) connected component (GCC) when a fraction of nodes is removed [11,17]. Figure 2 compares the robustness of the London Rail network (LRN) with equivalent Erdos-Renyi (Random) and Scale-Free (SF) network representations, subjected to random and targeted failure scenarios (see Methods: Failure Scenarios). Equivalence of the real (LRN) and simulated (Random and SF) networks in this context implies that the total number of nodes, layers and average degree are kept identical. Initially, our analysis treats the inter-network links in the LRN as connectivity (enabling) links. For random failure, 20 independent runs are conducted where nodes are removed at random and the ensemble mean for the GCC is plotted. For targeted failure, two of most common centrality measures – node degree and betweenness – are considered. Similar failures are applied to the synthetic Erdos-Renyi (ER) and Scale-Free networks and results are shown in Figure 2 for (A) random failure, (B) targeted (degree) and (C) targeted (betweenness).

When inter-network links are treated as connectivity links, the overall network effectively behaves like a supra-single network with distinct communities. Thus, a question from the network science perspective that may arise is why represent the LRN as a NoN in the first place. Our rationale for the proposed NoN representation is that by increasing topological granularity (i.e., dividing the NoN into interconnected networks) new insights can be derived which otherwise would not be possible. Thus, insights about the *interdependence* of the different network layers and the ensuing *cascading failure* may be better developed through a NoN representation. Furthermore, this provides a general framework for interdependent critical infrastructure networks such as the power grid and communication networks. Here the LRN NoN representation allows an examination of the interdependency between the three different yet coupled rail networks (Underground, Overground, DLR).

Based on the consideration above, the next part of our analysis treats the inter-network links as dependency links, i.e., if a node fails in one layer, then its dependent node in the corresponding dependent layer fails as well. Without loss of generality, in our work, we consider the two largest layers, specifically, Underground and Overground Rail networks, and study the system by removing nodes only from the Underground layer. As Underground nodes are removed, our representation removes the dependent nodes in the Overground layer, which in turn provides a feedback effect on the Underground network. Thus, the feedback between the layers leads to a cascading failure scenario. Supplementary Figure S6 presents a schematic diagram to illustrate this cascading failure scenario. Corresponding results for this scenario are shown in Figure 2 (D) for random failure, (E) targeted failure (based on degree) and (F) targeted (betweenness).

The LRN is found to be the least robust during random failure for both single-aggregate and NoN representations. However, when targeted failure is considered, the LRN is found to be less robust than Random but more robust than SF networks. The extreme vulnerability of SF networks against targeted attacks is well-established [21]. The relatively high vulnerability of real-world transportation network like the LRN can be attributed to the sparsity of connections and a narrow degree distribution concentrated around ~2 (Supplementary Figure S7), resulting in the presence of a few highly critical nodes (junction stations). Moreover, the topology can be described as 2D lattice-like and spatially constrained [22], i.e. nodes are linked mostly to nearby nodes as dictated by efficient design considerations (since creating each physical link has an associated cost). Consequently, while spatial networks have short links, random networks do not have a characteristic link length because nodes can be spatially mobile, and links can be across the networks thus leading to a compact topology and higher robustness.

Furthermore, the average shortest path length – a commonly used metric in network science - is significantly higher in the LRN (Table 1), which signifies that a select few nodes are extremely critical through which most of the shortest paths pass (i.e. there is an absence of hops). These are usually the transit points in transportation networks, and their failure may lead to significant loss of network functionality in quick time. On the contrary, the LRN robustness is boosted by the fact that degree correlation between dependent nodes is one (since dependent nodes are identical in this case). Indeed, it has been shown that high inter-similarity in spatial NoNs makes them more robust [23].

For a more comprehensive understanding, we studied the role of connectivity and dependency strength on the robustness of NoNs in general. We consider two equal sized ER random networks under random failure. Dependency ratio 'p' is defined as the fraction of nodes interdependent in each network. For different values of p, robustness profiles are plotted (Figure 2(G)). We find that as dependency increases, the coupled network fails faster. However, below a certain dependency ratio p, the coupled network doesn't fail completely, as can be seen in the plot where the final GCC size never approaches zero. For higher values of p, not only do we observe a complete failure, but also a transition from a second order to first-order (abrupt) failure profile, which is in conformation with existing literature [24]. To understand the role of connectivity, keeping p=1, the average degree <k> is varied in both networks (Figure 2(H)). It is found that coupled networks with higher average degree are more robust. In other words, our findings suggest that increased redundancy has a positive impact on network robustness.

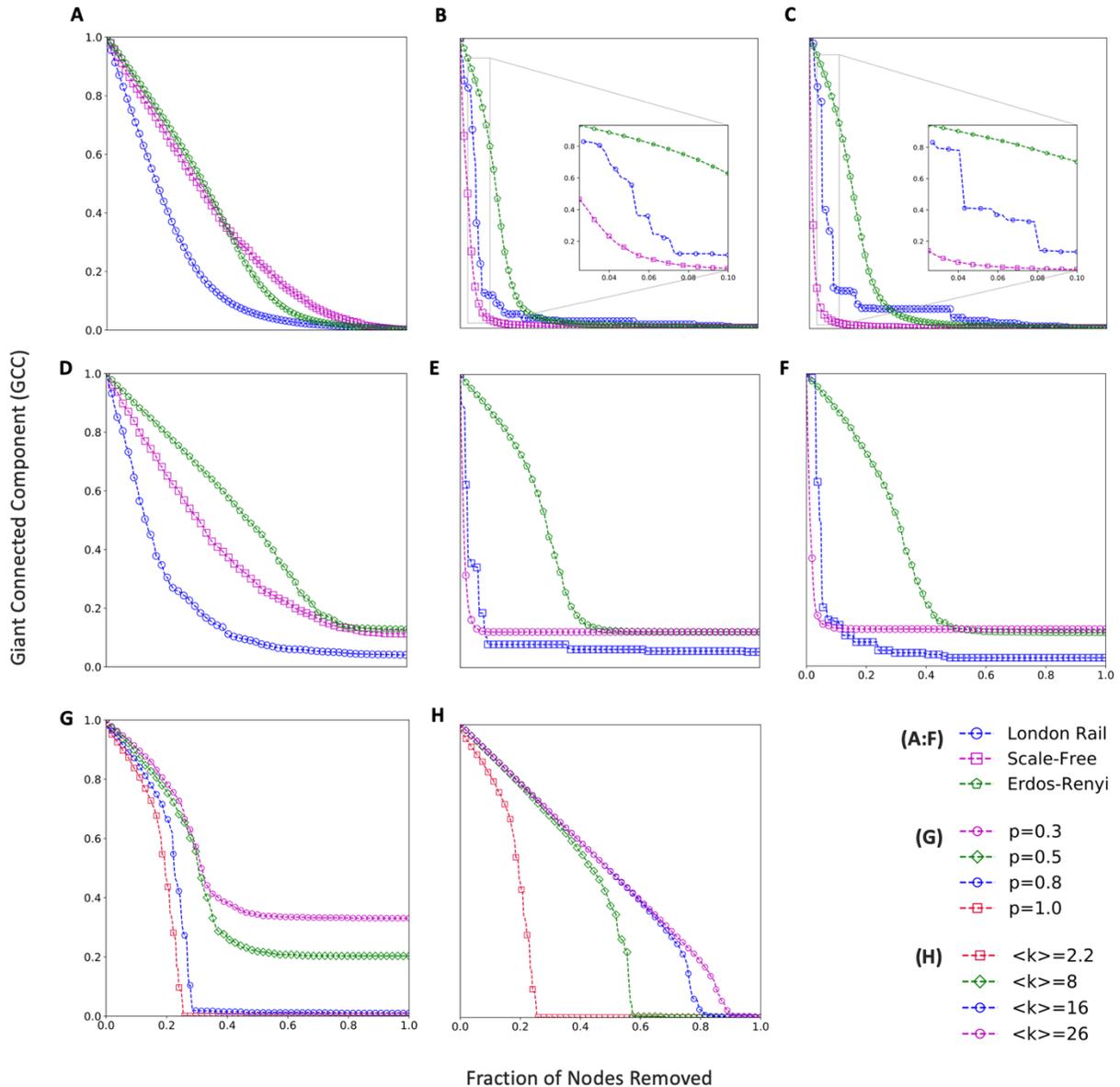

**Figure 2 | Robustness of London Rail Network.** The size of the largest (giant) connected component is plotted as nodes are removed from the network. Simulated Erdos-Renyi (Random) and Scale-Free networks are used as benchmark. In figure labels (A, B and C) the network is treated as aggregated (single-layer) and in (D, E and F) as a multilayer network-of-

networks (NoN). Nodes are removed randomly in **(A) and (D)**; in a targeted manner based on degree (**(B) and (D)**) and betweenness (**(C) and (E)**). To simulate cascade failure in the NoN case **((D) to (F))**, nodes are removed only from the underground network. **(G) and (H)** Impact of interdependency (coupling strength) and average network degree, respectively, on robustness demonstrated using two coupled Erdos-Renyi networks. Each plot is an ensemble mean of 20 independent runs.

*Table 1: Average Shortest Path Length*

| Network Type | Average Shortest Path Length |
|---|---|
| **London Rail Network** | ~13 |
| **Erdos-Renyi Random Network** | ~4.8 |
| **Scale-Free Network** | ~6.1 |

## 2.2 Flood Failure – Robustness and Recovery
### 2.2.1 Identifying Flooded Nodes
The London Rail NoN is geocoded over the map of London using the lat-long coordinates of the stations (Figure 1). Next, the 100-1000-year flood risk map on river Thames, obtained from the UK's Environment Agency's database [25], is overlaid on this network. Figure 1(A) shows the impacted stations (nodes) lying in the flood risk zone - 65 stations in total are impacted.

After the flooded nodes are identified, we test the LRN against different failure scenarios – flood-induced, random, local, targeted and compound. Their mechanism are described in the Methods section. Similar to before, we use the giant connected component (GCC) to quantify the robustness of the network. We find that flood induced failure has a distinct profile compared to both the types of random failures (Figure 3(A)). Approximately 80% of the total damage happens within the first one-third node failures in the case of flooding. The kink (sharp drop in slope) in the flood-induced failure curve is indicative of the potential for sudden large-scale failure of the network as well as the existence of a critical point. The reason for sudden failure in flooding is that the average degree of the flooded nodes is higher than the average degree of the overall network (Supplementary Figure S8). In other words, more critical or important nodes are in the flood risk zone. For instance, in the case of the London subway, the busiest stations like King's Cross and Waterloo are next to river Thames and in the riparian zone. Historically, cities have developed around rivers and the important stations and junctions were traditionally built close to the river body to provide quick transition to water transportation. However, the same design philosophy, which may not be as valid in modern times, leaves these transportation systems highly vulnerable to flooding.

### 2.2.2 Compound Disruptions – Flood and Targeted Attacks

Next, we consider the potential loss of functionality due to a compound disruption, as exemplified by opportunistic targeted attacks, specifically, where an adversary attempts to take advantage of reduced system functionality (caused by flooding) to create further damage through targeted attacks on the network. To model a compound failure, we remove n% of nodes in a targeted manner (based on degree centrality) after a certain proportion of flood has occurred. To have a fair comparison the total number of failed nodes (flood + targeted) are kept equal to the case of flood-only failure. In Figure 3(B), we find that even a targeted removal of 5% of total nodes in conjunction with flooding can cause network-wide failure, comparable to a purely targeted failure. The latter may be prohibitively difficult to execute. As the network is already fragmented and capacity is weakened, it only takes a select few nodes more to cause a disproportionately large damage.

### 2.2.3 Recovery Strategies

Quantifying resilience entails measuring not just the failure but also the recovery processes. In Figure 3(C), we compare recovery strategies based on different centrality measures. Random and reverse-order (node farthest from the river first) recovery of flooded nodes are plotted as baselines. Recovery is fastest when nodes are added back in order based on degree and betweenness centrality. When a node with high betweenness is added back, it re-enables a greater number of (shortest) paths which pass through it. Understandably, given the importance of path lengths, this property is most critical in a transportation network, as discussed in the literature [26, 27].

However, in the interests of generality, it is useful to mention here that for other infrastructure systems recovery based on degree (or other network attributes) may be faster. Here we also find that not all centrality measures necessarily will yield faster recoveries, such as the eigenvector centrality in this case. As can be seen in Figure 3, recovery based on eigenvector centrality is not much better than random recovery for the LRN. Thus, centrality measures which quantify the criticality of the network nodes may not always conform to the critical nodes in real-world network. Centrality measures are only one way of achieving a faster recovery in a network, however, decision-making in real-world systems is more complex with conflicting cost and resource constraints at play. Thus, while centrality measures or other network attributes may offer preliminary guidance and perhaps an organizing principle to the recovery process, there is a need to explore recovery in depth in future research.

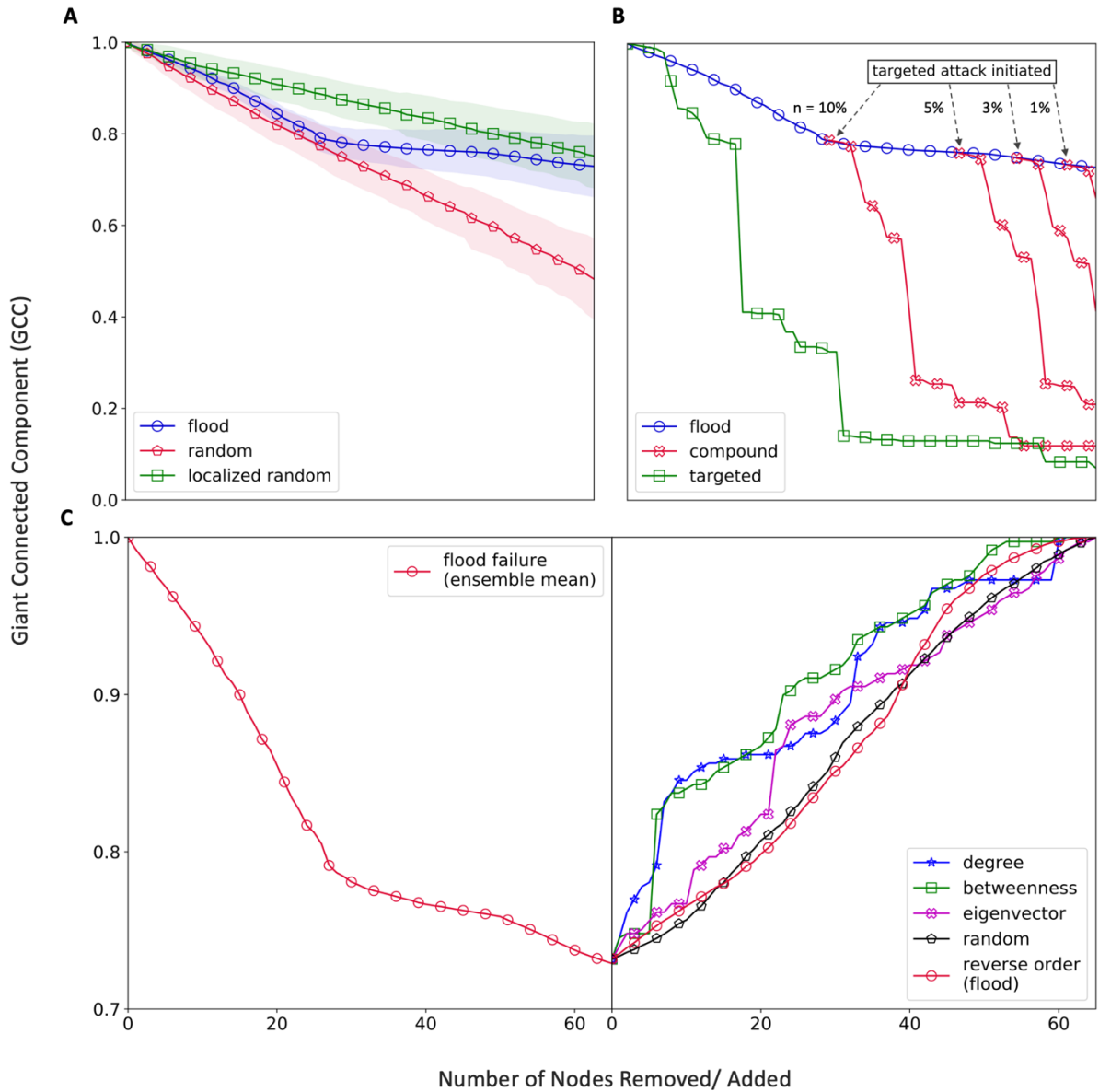

**Figure 3 | Flood and Compound Failure – Robustness and Recovery.** Robustness of the London Rail network under **(A)** flood failure and two types of random failure (overall and localized). **(B)** Compound failure – where n% nodes (of total) are removed in a targeted manner after certain nodes have failed due to flooding. **(C)** Recovery strategies post flood-failure. Nodes are added back to the network in an order derived from different centrality measures. Random recovery and reverse-ordered recovery of flooded nodes are also plotted for comparison. [In case of flood, random and local failure], each plot is an ensemble mean of 20 independent runs.

## 2.3 Realistic Scenario – Flooding of the Underground Network

Taking cue from a real incident in 2012 when Hurricane Sandy inundated a major section of the New York subway (e.g., see Figure 2 in [28]), we model a similar disruption in the LRN. Here, we remove the flooded nodes only from the Underground layer. Due to inter-network dependency, failure will be propagated to the other two layers.

Figure 4 shows how indirect failures caused by cascading failures can be disproportionately severe in the dependent layers, as exemplified here in the LRN. This is because the same lower degree nodes in one network can be more critical (with higher degree) in the dependent networks and cause greater disruption, an effect that may multiply when multiple such nodes and networks are considered. Figure 5 presents a schematic illustration of this process.

For completeness, we model a compound-failure case as in the one above. We observe that a 5% targeted attack can amplify the network damage by multiple times (over 300% in the case of LRN Underground) and may even cause a complete failure in a dependent network, such as the DLR in this case (Figure 4(B)).

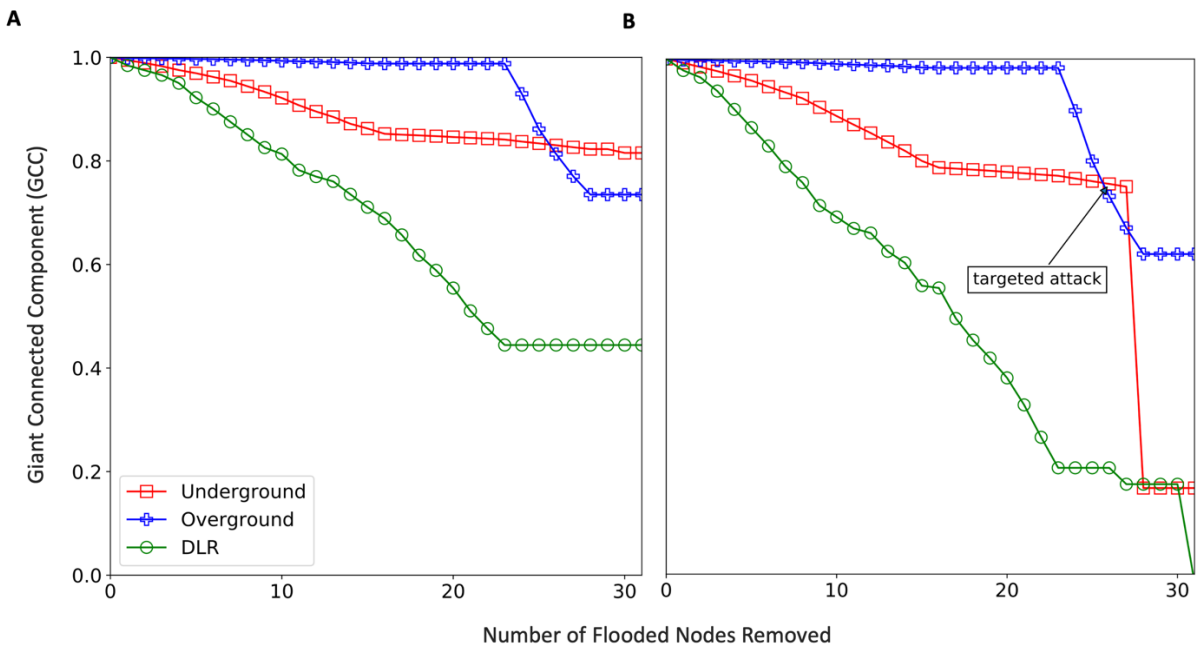

**Figure 4 | Indirect Flood Failures due to Interdependency**.
Flood failure is restricted to the underground network and indirect failure in the other two networks is analyzed. **(A)** Flood-only failure **(B)** Compound failure – same as before. As observed, indirect failure can be disproportionate due to different criticality of the same node across networks.

The results above strengthen our hypothesis that compound failures are disproportionately more damaging when compared to individual attacks. In fact, the whole (i.e., damage from compound extremes) may be greater than the sum of the parts (i.e., the sum of damages from the individual extremes occurring separately). However, the latter hypothesis would need to be further examined in future studies. A generalization of these insights to urban transportation NoNs would require examining more urban transportation systems, while a broader generalization would need critical interdependent networks spanning the transportation, power, communications and water sectors.

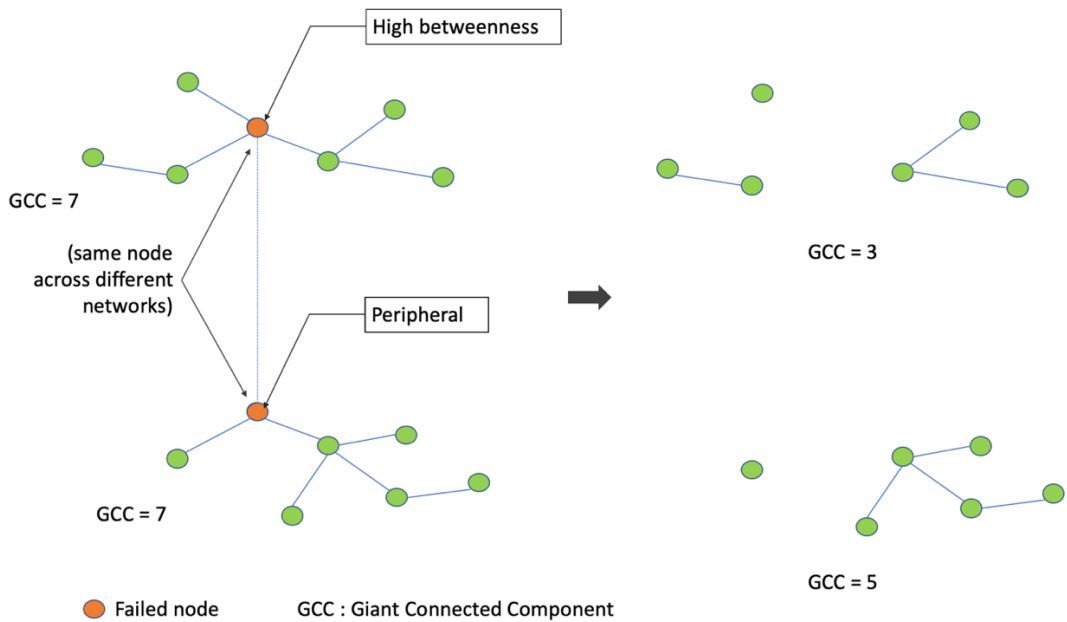

**Figure 5 | Node Criticality and Severity of Indirect Failure**
Schematic figure to demonstrate extent of indirect failure given failed node with different criticality (e.g. betweenness) across interdependent network. Node in orange is the failing node and GCC is size of the giant (largest) connected component. A higher node criticality translates to greater loss of network functionality (lower size of final GCC).

**Discussion**
As urbanization and climate change intensifies, understanding and improving the resilience paradigm of urban lifeline infrastructure is critical. Here we have presented an analytical framework to quantify the resilience of multimodal urban transport systems (MUTS) modeled as network-of-networks (NoNs), which can be viewed as extension of the recent works on resilience of single transportation networks. [9, 33]. We also show how these spatial network-of-networks are extremely vulnerable to a suite of threats including natural hazards, with a special focus on compound threats (e.g. cyber or cyber-physical attacks during flooding) which may cause disproportionate and network-wide failures.

Based on our findings here and insights in the literature [29, 30], we suggest that urban transportation systems, being a type of spatially embedded networks, have intrinsic topological vulnerabilities that are considerably different from their non-spatial counterparts. Thus, theoretical frameworks (such as [14, 17-19]) in the extant network science literature which may have been developed for non-spatial networks, may not directly translate to the real-world without incorporating the spatial nature of these infrastructure networks. This finding points to the need for broadening network science studies in the context of real-world infrastructure resilience by developing theoretical and empirical studies, as well as methods and tools, that are geared toward spatially embedded networks.

Furthermore, urban rail systems in long-established megacities (especially in the developed nations) such as London or New York are more than a century old when the idea of system-level resilience may not have been the highest priority. Our results here show how design decisions which maximized efficiency and relatively obsolete considerations (e.g., the need for major stations to be near water bodies) from pre-modern times rendered the London Rail Network highly vulnerable to flood-induced failure. Thus, there is a need for designers of new urban lifeline infrastructure networks (including transportation), as well as for owners and operators of existing urban lifeline systems, to embed resilience considerations and move beyond traditional design practices which focus only on maximizing efficiency and structural longevity. This is especially true under growing threats of urban floods and other natural hazards, as well as compound threats such as opportunistic cyber or cyber-physical attacks. Future research needs to strengthen these insights by considering multiple urban systems across geographies and lifeline sectors.

Our study further suggests the potential of developing novel actionable insights on resilience that may be used by practitioners and stakeholders, including urban infrastructure owners and operators. Thus, based on our findings, we can state that flooded nodes (stations) with lower degrees (connectivity), i.e., those that form a connecting link between two dense clusters, can cause an immediate breakdown of an interdependent network-of-networks into sub-components or even communities of sub-networks. This finding is non-intuitive, in view of the common understanding [31, 32] in hazard mitigation that there is a need to focus on nodes considered conventionally 'important' (e.g., in terms of higher degree). We suggest the need for broader research in networks-based analysis, which we believe (based on our findings here and on the literature [33-35]) can provide valuable information to stakeholders such as how resilience can be baked-in (for new design) or enhanced (for existing systems) by appropriately modifying network topology, making infrastructure systems more modular, and through broader redistribution of critical dependencies.

A second advancement vis-à-vis [9, 14, 33] is that we consider the emerging threat of compound disruptions, especially targeted attacks in conjunction with natural hazards. We find that when the network capacity is already weakened, a limited targeted attack is sufficient to cause disproportionate damages and network-wide failures. However, our work is an early exploration of networked systems under compound attacks. A hypothesis-driven study which explores this topic in more depth both from the perspective of network science theory and engineering practice appears to be a clear and present need.

Our findings and frameworks lead to potentially new hypotheses and research directions, some of which have been described earlier in this section. As a next step, there is a need to incorporate system dynamics and flow in addition to topology for NoNs resilience, especially for spatially embedded NoNs. In addition, the giant component size (GCC) as a robustness measure may be more suited for non-spatial networks, thus there is need to explore alternative metrics which consider network dynamics. Finally, considering multiple failure schemes (including but not limited to cascading failure schemes considered here) and contingency analysis of networked systems (including unknown or potentially unknowable threats) may yield insights of value to both researchers and practitioners.

**Methods**

**Dataset Availability**

The London metro dataset used in the paper has been made available by De Domenico et al. [36]. The flood risk map is obtained from the UK's Environment Agency database. To facilitate reproduction of results, all Python codes are available on the lead author's GitHub account (github.com/nisyad).

**Network Structure**

| Network Layer | Nodes (Stations) | Links |
| --- | --- | --- |
| Underground | 271 | 312 |
| Overground | 83 | 83 |
| DLR | 45 | 46 |

**Common Stations**

There are 24 stations common between the Underground and Overground networks, 5 stations common between the Underground and DLR, and 1 station is common between the Overground and DLR.

**Failure Scenarios**

**Flood Failure**: The flooded nodes are divided into three categories based on their proximity to river Thames. Considering the outward spread of the flood, the first category of nodes closest to the river fail first and so on. To account for uncertainty, nodes within each category may fail randomly. 20 independent runs are considered to model flood failure.

**Random Failure**: Nodes are removed randomly from the overall network.

**Random-Local Failure**: Using the lat-lon values of station locations, a distance matrix is generated for the entire network where each entry (i, j) is the distance between station i and j. 100 local clusters of geographically closest 65 nodes (same number as flooded nodes) are generated across the network to model a localized failure. Within each cluster, nodes fail at random. 20 independent runs considered.

**Targeted Failure**: To model a targeted failure, a predetermined order of nodes based on some metric of importance is considered and nodes are removed in the decreasing order of importance. Here the importance metric used are degree and betweenness centrality.

**Compound Failure**: After a certain level of flood failure has occurred, a targeted attack is introduced by removing n% of nodes based on degree centrality. The overall number of failed nodes are kept equal to the number of total flooded nodes.

**Cascading Failure: Algorithm**

1. **INPUT**: network A, B, nodes_to_remove, dependent_nodes
2. INIT Empty List GCC
3. **FOR** node in nodes_to_remove
4.    GCC List <- append(GCC size)
5.    REMOVE: node in network A
6.    REMOVE: nodes in network A not part
7.        of GCC
8.    INIT new_nodes_to_remove := nodes
9.       removed in (5) and (6) if present in list of
10.       dependent node
11.    REVERSE: network A and B
12.    nodes_to_remove := new_nodes_to_remove
13.    RECURSIVE CALL: to (4)
14.       REMOVE: dependent nodes in network B
15.       REMOVE: nodes in network B not part
16.          of GC
17.    STOP: no more nodes to remove
18. **END LOOP**
19. **RETURN**: List of GCC

† GCC = giant connected component

```
Compound Failure: Algorithm

    1.  INPUT: network G, flooded_nodes_to_remove, n%
        targeted attack, targeted_nodes_to_remove
    2.  INIT Empty List GCC
    3.  IF Cascade Failure IS TRUE:
    4.     CASCADE FAILURE (Network G,
    5.                      flooded_nodes_to_remove)
    6.     CASCADE FAILURE (Network G, n,
    7.                      targeted_nodes_to_remove)
    8.  ELSE
    9.     FOR node in flooded_nodes_to_remove
    10.    REMOVE: node in network G
    11.    GCC List <- append (GCC size)
    12.    STOP: no more nodes to remove
    13.    END LOOP
    14.    RETURN: List of GCC
```

**Centrality Measures**

Complex network structures are very heterogenous and some nodes are expected to be more important than other nodes which can be quantified by node centrality measures.

**Degree Centrality:** number of nodes linked to a particular node.

**Betweenness Centrality**: for every pair of nodes, there exists a shortest path. The betweenness centrality of a node is number of such of shortest paths passing through it.

$$g(v) = \sum_{s \neq v \neq t} \frac{\sigma_{st}(v)}{\sigma_{st}}$$

$\sigma_{st}$ is the total number of shortest paths from node $s$ to $t$ and $\sigma_{st}(v)$ is the total number of shortest paths passing through node $v$.

**Eigenvector Centrality**: a node can be considered important if it is connected to other important nodes. This importance of node $i$ can be quantified by a vector $x_i$:

$$x_i = \frac{1}{\lambda} \sum_{k=1}^{N} A_{k,i} x_k$$

where $\lambda$ is a non-zero constant. In the matrix form:

$$\lambda x = Ax$$

The importance of node $i$ is defined by the left-hand eigenvector of the adjacency matrix $A$ associated with eigenvalue $\lambda$. The entries of $x$ are called eigenvector centrality.

**Acknowledgement**

Funding for this work was provided by the Pacific Northwest National Laboratory's (PNNL) National Security Directorate Mission Seed Laboratory Directed Research and Development (LDRD) Program, as well as the US National Science Foundation's BIGDATA Award (1447587), INQUIRE Award (1735505) and CyberSEES Award (1442728). PNNL is a multiprogram national laboratory operated by Battelle for the United States Department of Energy under DE-AC05-76RL01830. A significant portion of this work was conducted at the PNNL campus in Richland, WA, as part of the 2019 National Security Internship Program (NSIP). Yadav defined the problem, performed the analysis, interpreted the results, and wrote the paper, while Chatterjee and Ganguly defined the problem, helped guide the analysis, interpreted the results, and wrote the paper. The authors are grateful to the CoMuNe Lab (comunelab.fbk.eu/data.php) for providing open access to the London Rail Network data, as well as to Elizabeth Mary Warner and Craig Poulin of Northeastern University and Domenic Skurka of the Pacific Northwest National Laboratory for helpful discussions. The authors have no competing interest to declare.

**Supplementary Figures**

**Figure S6** | Schematic Cascade Failure in Partially Interdependent Network. In stage I, a node is removed due to direct failure (orange color). In stage II, all the dependent nodes in both layers are removed (yellow color). In stage III, dependent nodes on the removed node in the other layer are removed. Stage IV represents a feedback mechanism where dependent nodes (in first layer) on the failed nodes in the other layer are also removed (purple color). Green nodes represent functional nodes in each stage as part of the giant connected component.

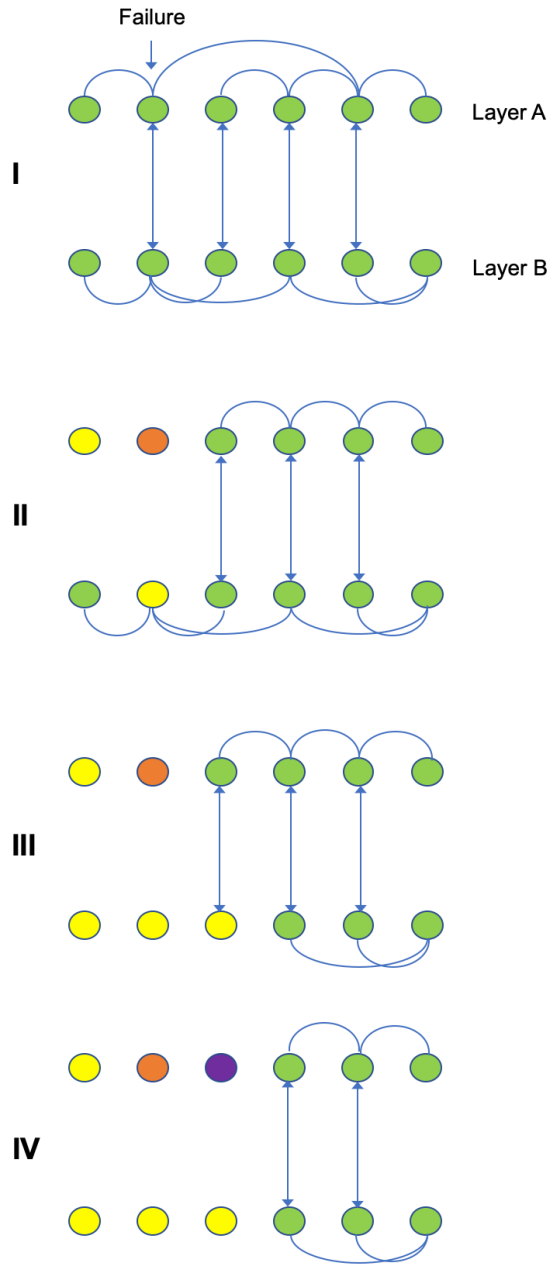

**Figure S7** | Degree Distribution of the London Rail Network. Equivalent Erdos-Renyi and Scale-Free (SF) network degree distributions are plotted for comparison. Degree-distribution of the scale free network is curtailed at 14 in the plot.

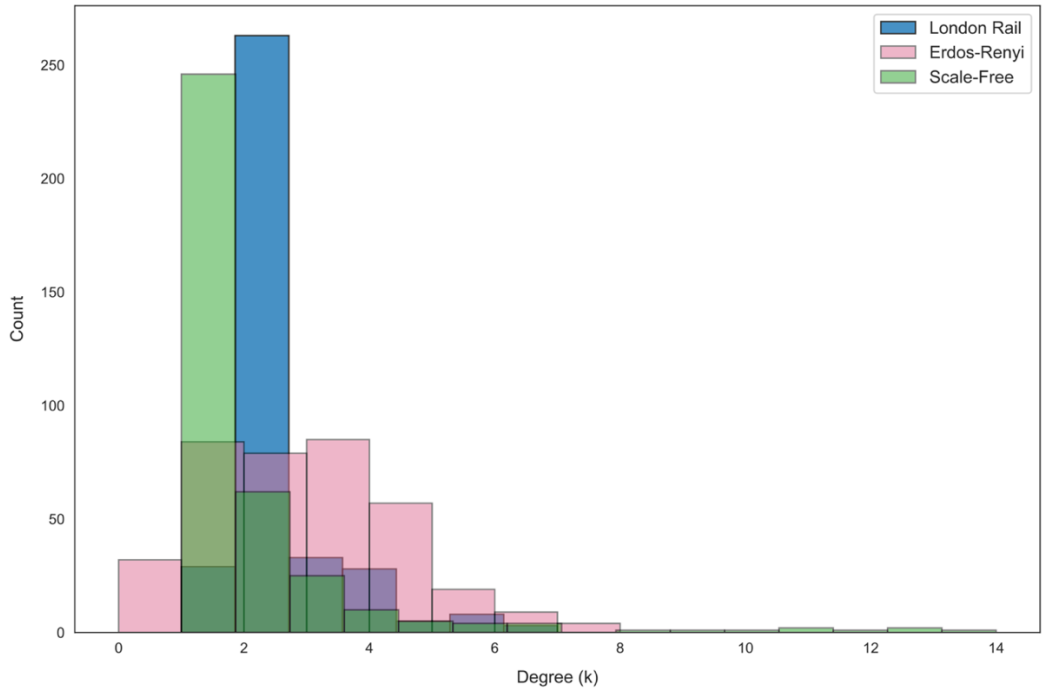

**Figure S8** | Degree Distribution of the Flooded Network vs the Total Network. Higher average degree for the flooded network indicates the presence of more critical nodes in the flooded area.

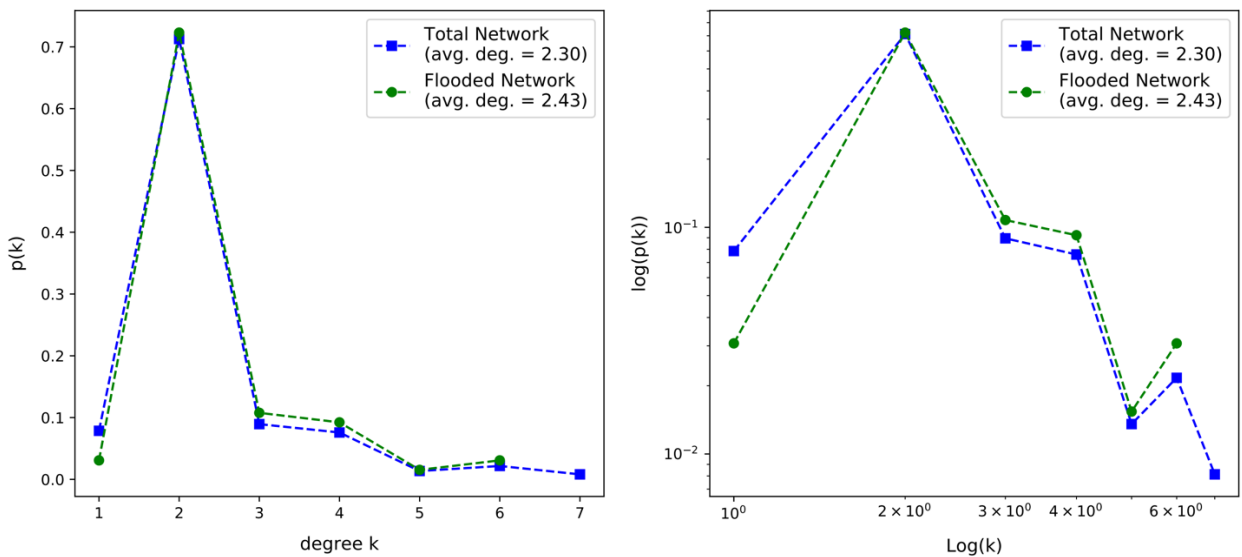